\documentclass[12pt,preprint]{aastex}
\usepackage{amssymb}
\usepackage{epsfig}
\usepackage{natbib}
\usepackage{graphicx}
\usepackage{color}
\usepackage{tabularx}
\usepackage{epsfig}
\usepackage{rotating}
\usepackage{multirow}
\usepackage{subfigure}
\usepackage{ulem}
\usepackage{longtable}

\newcommand       \mum        {\,{\rm \mu m}}

\newcommand       \Ks           {{K_{\rm s}}}

\newcommand{\AJ}{A_{\rm J}}
\newcommand{\AH}{A_{\rm H}}
\newcommand{\AKs}{A_{\rm K_S}}
\newcommand{\AV}{A_{\rm V}}

\newcommand{\RV}{R_{V}}
\newcommand{\Rv}{R_{V}}
\newcommand{\RI}{R_{I}}

\shorttitle{Cepheid distances and mid-infrared period--luminosity relations}
\shortauthors{S. Wang et al.}

\begin{document}

\title{
The Near-infrared Optimal Distances Method Applied to Galactic
Classical Cepheids Tightly Constrains Mid-infrared Period--Luminosity
Relations }

\author{Shu Wang\altaffilmark{1},
             Xiaodian Chen\altaffilmark{2}, 
             Richard de Grijs\altaffilmark{1,3,4}, 
             and Licai Deng\altaffilmark{2}}
\altaffiltext{1}{Kavli Institute for Astronomy and Astrophysics,
                 Peking University,
                 Beijing 100871, China;
                 {\sf shuwang@pku.edu.cn}
                 }
\altaffiltext{2}{Key Laboratory for Optical Astronomy, 
                 National Astronomical Observatories, 
                 Chinese Academy of Sciences, 
                 Beijing 100012, China; 
                 {\sf chenxiaodian@nao.cas.cn} 
                 }
\altaffiltext{3}{Department of Astronomy,
                 Peking University,
                 Beijing 100871, China
                 }
\altaffiltext{4}{International Space Science Institute--Beijing,
                 Beijing 100190, China
                 }

\begin{abstract}
Classical Cepheids are well-known and widely used distance indicators.
As distance and extinction are usually degenerate, it is important
to develop suitable methods to robustly anchor the distance
scale. Here, we introduce a near-infrared (near-IR) optimal distance
method to determine both the extinction values of and distances to a
large sample of 288 Galactic classical Cepheids.
The overall uncertainty in the derived distances is less than 4.9\%.
We compare our newly determined distances to the Cepheids in our
sample with previously published distances to the same Cepheids with
{\sl Hubble Space Telescope} parallax measurements and distances based
on the IR surface brightness method, Wesenheit functions, and the
main-sequence fitting method. The systematic deviations in the
distances determined here with respect to those of previous
publications is less than 1--2\%. Hence, we constructed Galactic mid-IR
period--luminosity (PL) relations for classical Cepheids in the four
{\sl Wide-Field Infrared Survey Explorer} ({\sl WISE}) bands ($W$1, $W$2,
$W$3, and $W$4) and the four {\sl Spitzer Space Telescope} bands
([3.6], [4.5], [5.8], and [8.0]). Based on our sample of hundreds of
Cepheids, the {\sl WISE} PL relations have been determined for the
first time; their dispersion is approximately 0.10 mag. Using the
currently most complete sample, our {\sl Spitzer} PL relations
represent a significant improvement in accuracy, especially in the
[3.6] band which has the smallest dispersion (0.066 mag). In addition,
the average mid-IR extinction curve for Cepheids has been obtained:
$A_{W1}/A_\Ks \approx 0.560$, $A_{W2}/A_\Ks \approx 0.479$,
$A_{W3}/A_\Ks \approx 0.507$, $A_{W4}/A_\Ks \approx 0.406$,
$A_{[3.6]}/A_\Ks \approx 0.481$, $A_{[4.5]}/A_\Ks \approx 0.469$,
$A_{[5.8]}/A_\Ks \approx 0.427$, and $A_{[8.0]}/A_\Ks \approx 0.427$
mag.
\end{abstract}
\keywords{distance scale --- stars: variables: Cepheids --- infrared:
  ISM --- ISM: extinction}

\section{Introduction}\label{intro}

Classical Cepheids are well-known and widely used distance indicators
in relation to the well-established period--luminosity (PL) relation
(the `Leavitt law'; Leavitt \& Pickering 1912). By employing the PL
relation, Cepheids can be used to measure nearby extragalactic
distances, constrain the Hubble constant, and study Galactic structure
and kinematics. Because Cepheids are sparsely distributed throughout the
Galaxy, they suffer from distinct reddening effects for each
sightline. Their apparent magnitudes are therefore generally reddened
and attenuated by intervening interstellar dust to varying extents
(Madore et al.\ 2017). Determining the empirical PL relations for
Galactic Cepheids requires measuring their distances and corrections
for the wavelength-dependent extinction effects pertaining to
individual Cepheids.  However, distance and extinction are often
tightly coupled.

Before determining Cepheid distances, the wavelength-dependent
extinction values need to be measured for individual Cepheids. To
reduce the influence of extinction, multi-band photometry is usually
employed to obtain reddening-free magnitudes, such as the widely used
Weisenheit functions, $W = V-\Rv(B-V)$ and $W = V-R_I(V-I)$
(Madore\ 1976, 1982; Madore \& Freedman\ 1991; Fouqu\'e et al.\ 2007;
Turner 2010). For a constant value of $\Rv$ or $\RI$, Weisenheit
magnitudes can be derived directly for Cepheid distance
measurements. However, adopting constant $\Rv$, $\RI$ values means
that we implicitly assume that the optical reddening law is
universal. Yet, the optical extinction law, usually expressed as
$A_\lambda/\AV$ at $\lambda < 0.7\mum$, is known to vary significantly
among sightlines (Cardelli et al.\ 1989; hereafter CCM). CCM found
that the variation can be described by the optical total-to-selective
extinction ratio $\RV = \AV /E(B - V)$. The average extinction law for
diffuse, low-density regions in the Galaxy is $\RV =3.1$, which is
commonly used to correct observations for dust extinction. In fact,
the optical extinction law exhibits significant diversity even within
small regions (in angular size) of the diffuse interstellar medium
(Wang et al.\ 2017). Therefore, Weisenheit functions cannot be applied
to dense environments. 

The influence of extinction in near-infrared (IR) bands is much less
than that in optical bands, e.g., $\AJ/\AV = 0.29$, $\AKs/\AV = 0.12$
for the average extinction law of Galactic diffuse regions, adopting
$\Rv=3.1$. The near-IR ratio of total-to-selective extinction, such as
$\AKs/E(H-\Ks)$ or $\AJ/E(J-\Ks)$, has been adopted to determine the
near-IR extinction and distances to Cepheids. However, the
$\AKs/E(H-\Ks)$ values still show variations. The interstellar
extinction law toward the Galactic center determined by Nishiyama et
al.\ (2006) is $\AKs/E(H-\Ks) = 1.44\pm0.01$; Nishiyama et al.\ (2009)
determined $\AKs/E(H-\Ks) = 1.61\pm0.04$. This will cause at least a
10\% distance uncertainty, which is a key problem in studying the
structure of the Galactic bulge.

With independent access to Cepheid distances, the empirical PL
relation can be determined directly by fitting the period versus
absolute magnitude trends in different filters. In the last century,
optical $BVI$-band photometry was usually used to constrain the PL
relation (Madore \& Freedman 1991; Gieren et al.\ 1998; Tammann et
al.\ 2003; and references therein). However, the empirical PL relation
has an intrinsic dispersion which is caused by the finite width of the
instability strip for pulsating stars. This dispersion is particularly
significant in optical bands (e.g., in the $B$ filter is amounts to
$\sim0.2$ mag), but it decreases toward longer wavelengths (e.g., in
near-IR and mid-IR band, the dispersion is $< \sim0.1$ mag; see, e.g.,
Madore \& Freedman 1991; Inno et al.\ 2013; Gaia Collaboration 2017;
and references therein). Therefore, compared with the optical bands,
the PL relations in near-IR bands exhibit less dispersion and fewer
systematic errors (Madore \& Freedman 1991; Gieren et al.\ 1998;
Fouqu\'e et al.\ 2007; Monson \& Pierce 2011). In the past decade,
with the wealth of available near-IR photometry for Galactic Cepheids,
there have been major improvements in constraining the near-IR Cepheid
PL relations (An et al.\ 2007; Fouqu\'e et al.\ 2007; Monson \& Pierce
2011; Chen et al.\ 2015, 2017). More recently, and considering that
the effects of dust extinction in mid-IR bands are less significant
than in near-IR bands, we have seen an increase in interest in mid-IR
PL relations (Marengo et al.\ 2010; Monson et al.\ 2012; Ngeow\ 2012).
However, more Cepheids samples with accurate distances are needed to
reduce the remaining uncertainties in the Galactic mid-IR PL
relations.

In this paper, we have collected a large sample of 
Galactic classical Cepheids with IR data. 
Their distances have been determined accurately by carefully 
revisiting the near-IR extinction (Section 2). 
Mid-infrared PL relations for these Cepheids in the four {\sl Spitzer} 
and four {\sl WISE} bands are also derived in Section 3. 
Comparisons of the PL relations and an assessment of 
the mid-IR extinction law are discussed in Section 4. 
We summarize our principal conclusions in Section 5.

\section{Distances to the Galactic Classical Cepheids}
\subsection{Sample and Method}
%

To accurately determine the distances to Galactic classical
Cepheids in near-IR bands, a sample of Galactic classical Cepheids
with near-IR $J, H, \Ks$-band mean magnitudes has been collected from
the literature. Van Leeuwen et al.\ (2007) published 229 Cepheids with
near-IR mean magnitudes in the South African Astronomical Observatory
(SAAO) system. A sample of Galactic Cepheids with individual
Baade--Wesselink distances was compiled from publications by Fouqu\'e
et al.\ (2007), Groenewegen (2008), Pedicelli et al.\ (2010), and
Storm et al.\ (2011). Monson \& Pierce (2011) provided near-IR
photometric measurements for 131 northern Galactic classical Cepheids.
Chen et al.\ (2017) used 31 open-cluster Cepheids to obtain $JH\Ks$
Galactic Cepheid PL relations. After removing duplicate sources, our
final sample comprises 288 classical Cepheids. $J, H, \Ks$-band mean
magnitudes in the SAAO and European Southern Observatory (ESO) systems
were converted to the Two Micron All Sky Survey (2MASS;
Skrutskie et al.\ 2006) system using the color transformation
equations given on the 2MASS website
\footnote{http://www.ipac.caltech.edu/2mass/releases/allsky/doc/sec6\_4b.html}. 
The objects' names, pulsation periods, and $J, H, \Ks$-band mean
magnitudes are summarized in the first five columns of Table 1.

\begin{table}[h!]
\begin{center}
\caption{\label{tab:dist} 2MASS and {\sl WISE} mean magnitudes and
  distances for our sample of 288 Galactic Cepheids\tablenotemark{a}}
\tiny
\vspace{0.1in}
\begin{tabular}{lccccccccc}
\hline \hline
Cepheid & log($P$) & $\langle J \rangle$  & $\langle H \rangle$  & $\langle \Ks \rangle$   
& $\langle W1 \rangle$   & $\langle W1 \rangle$   & $\langle W3 \rangle$   & $\langle W4 \rangle$   
& $\langle \mu_0 \rangle$\\ 
  & [d] & (mag) & (mag) & (mag) & (mag) & (mag) & (mag) & (mag) & (mag)\\
\hline
S VUL                 & 1.839 & 5.410(0.012) & 4.806(0.011) & 4.586(0.015) & 4.409(0.150) & 4.011(0.179) & 4.240(0.035) & 4.169(0.045) & 12.787$\pm$0.086\\     
GY SGE                & 1.714 & 5.530(0.012) & 4.827(0.011) & 4.546(0.015) & 4.368(0.098) & 3.987(0.151) & 4.259(0.026) & 4.186(0.039) & 12.222$\pm$0.082\\    
V1467 CYG             & 1.687 & 8.150(0.013) & 7.278(0.012) & 6.961(0.016) & 6.758(0.150) & 6.764(0.074) & 6.776(0.044) & 6.851(0.242) & 14.410$\pm$0.127\\    
SV VUL                & 1.655 & 4.552(0.012) & 4.051(0.011) & 3.905(0.016) & 3.850(0.141) & 3.660(0.114) & 3.888(0.058) & 3.868(0.033) & 11.620$\pm$0.096\\    
V0396 CYG             & 1.522 & 6.031(0.012) & 5.037(0.011) & 4.619(0.016) & 4.297(0.164) & 3.996(0.142) & 4.055(0.022) & 3.880(0.075) & 11.364$\pm$0.097\\ 
... & ... & ... & ... & ... & ... & ... & ... & ... & ...\\
\hline
\end{tabular}
\tablenotetext{a}{The entire table is available in the online journal. 
  A portion is shown here for guidance regarding its form and content.}
\end{center}
\normalsize
\end{table}

The $0.9\mum < \lambda < 3\mum$ near-IR extinction follows a power law
defined by $A_{\lambda}\propto{\lambda^{-\alpha}}$, with the index
$\alpha$ spanning a small range of $1.61 < \alpha < 1.80$ (Draine 2003). However, some of the most recently published values of $\alpha$
have tended to become systematically larger, even reaching $\alpha >
2.0$ (Wang \& Jiang\ 2014). The widely adopted extinction laws of CCM,
Rieke \& Lebofsky (1985), and Weingartner \& Draine (2001) are all
characterized by $\alpha=1.61$. A steep power law, $\alpha =1.99$
(Nishiyama et al.\ 2006), toward the Galactic Center is also commonly
used to correct for extinction in the heavily obscured Galactic bulge.
As our target Cepheids are located nearby in the Galactic plane, 
here we assume $A_{\lambda}\propto{\lambda^{-1.61}}$ ($\lambda: J, H, \Ks$). 
The discrepancy in Cepheid distances caused by adopting a larger
value of $\alpha$, $\alpha = 1.99$, will be discussed in Section 4.1.
For each given distance, $d$, the near-IR extinction is calculated as
$A_\lambda = m_\lambda - M_\lambda $ - 5 log $d$ + 5, where
$m_\lambda$ is the $\lambda$-band mean magnitude, and $M_\lambda$
represents the absolute magnitude (which can be derived from the
near-IR PL relations). We adopt the near-IR PL relations of Chen et
al.\ (2017). The discrepancies in Cepheid distances caused by adopting
different near-IR PL relations will also be discussed in Section 4.1.
In practice, steps of 0.1 pc are adopted for distances in the
  range 10 pc $\le d \le 15$ kpc. For a given $d$, $A_\lambda$ is
  calculated as $A_\lambda = m_\lambda - M_\lambda - (5 \log d - 5)$
  for $\lambda = J, H, \Ks$. The most reasonable distance $d$ results
  when the extinction values $\AJ$, $\AH$, and $\AKs$ can all be fitted 
  by the $\lambda^{-1.61}$ power law, in a minimum $\chi^2$ sense. 
We refer to this method as the near-IR optimal distances method.  
The accuracy of this method depends on the extinction in the $J, H, \Ks$ bands,
which is in essence similar to the construction of the Weisenheit
functions, although the latter depend only on two bands (e.g., $B, V$
or $V, I$).  Therefore, the distances and extinction thus derived are
expected to be more reliable.

\subsection{Distances and Errors}

Based on the near-IR optimal distances method described in Section 2.1, 
the distance moduli of our sample of 288 Galactic Cepheids have been derived.
The uncertainty in the distance modulus originates from a few
contributors, including the errors in (a) the observed magnitude, (b)
the absolute magnitude, and (c) the extinction. In addition, in the
application of our method, the $J, H, \Ks$ bands are used
simultaneously to determine optimal distances; this introduces
(d) a $\sim 2.0$\% statistical uncertainty. 
These uncertainties in the Cepheid distances are tabulated in Table 2.
The errors in (a), the observed magnitude, 
come from the photometric uncertainty. 
The average photometric uncertainty is $\sim$0.015 mag in $J$, 
$\sim$0.014 mag in $H$, and $\sim$0.016 mag in $\Ks$,
which contribute only uncertainties of 0.4\% in the distance moduli. 
Because we use the near-IR PL relations of Chen et al.\ (2017) 
to derive the $J, H, \Ks$ absolute magnitudes, 
we adopt for the uncertainties in the PL relations to be the errors in (b), 
these absolute magnitudes, resulting in $\sim3.4$\%
uncertainties in the derived distances. 
The extinction errors are $\sim2.2$\% and come from the uncertainty 
in the theoretical near-IR extinction law (for details, see Section 4.1).
Combining these errors, the uncertainties in the optimal distances can be derived. 
The overall average distance uncertainty, 
including both systematic and statistical errors, 
is about 4.9\%, with a standard deviation of 1.5\%. 
This is, in fact, an upper limit to the error; 
the actual distance accuracy is higher (see the discussion in Section 2.3). 
The values of our derived distance moduli and their uncertainties 
are tabulated in the last column of Table 1.

\begin{table}[h!]
\begin{center}
\caption{\label{tab:error} Error contributions to the
    uncertainties in the Cepheid distances}
\vspace{0.1in}
\begin{tabular}{lc}
\hline \hline
Contributor & Uncertainty (\%) \\
\hline
(a) Observed magnitude & 0.4\\
(b) Absolute magnitude & 3.4\\
(c) Extinction & 2.2\\
(d) Statistical uncertainty related to our method & 2.0\\
\hline
Total & 4.9\\
\hline\hline
\end{tabular}
\end{center}
\end{table}
\subsection{Comparison with Previously Published Results}
%

\begin{figure}[h!]
\centering
\vspace{-0.0in}
\includegraphics[angle=0,width=7.0in]{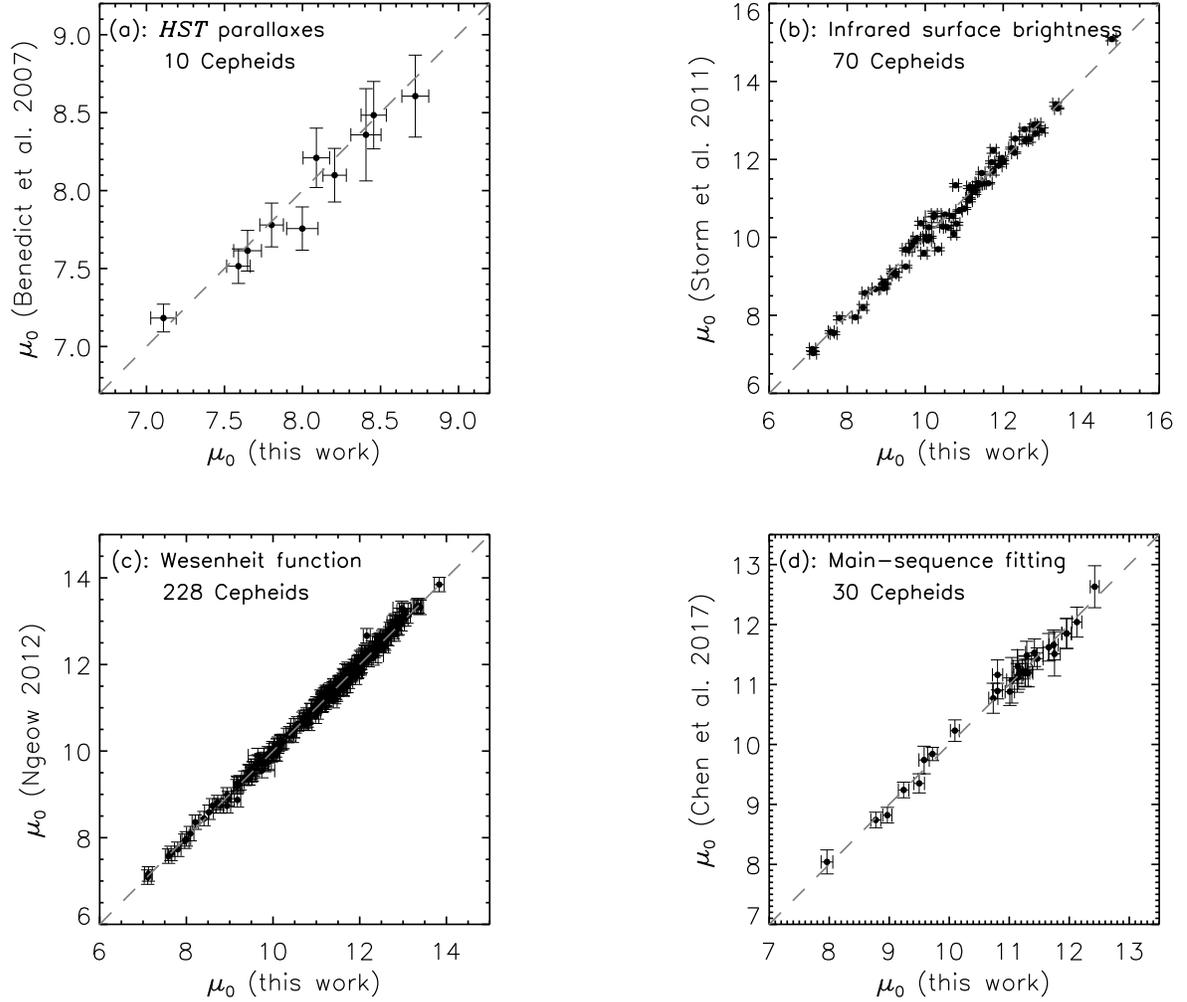}
\vspace{-0.2in}
\caption{\footnotesize
               \label{fig:dist_comp}
               Comparison of the distance moduli derived in Section
               2.1 (`this work') with distance moduli measured based
               on (a) {\sl HST} parallaxes (10 Cepheids: Benedict et
               al.\ 2007); (b) the IR surface brightness method (70
               Cepheids: Storm et al.\ 2011); (c) the Wesenheit
               function (228 Cepheids: Ngeow\ 2012); and (d)
               the main-sequence fitting method (30 Cepheids: Chen et
               al.\ 2017). The dashed lines are the $y = x$ loci and
               not fits to the data.}
\end{figure}

To analyze the systematic errors in our distances, we compare our
results with independently measured distances from the recent
literature; see Fig.~\ref{fig:dist_comp}. A quantitative comparison of
the systematic deviations is discussed separately.

We compared our distances with those obtained for the 10 Cepheids
that have {\sl Hubble Space Telescope} ({\sl HST}) parallaxes
(Benedict et al.\ 2007) shown in Fig.~\ref{fig:dist_comp} (a). 
The average distance difference between both of our methods is about 1.9\%.
The most recent Cepheid distances based on the IR surface brightness
method were published by Fouqu\'e et al. (2007), Groenewegen (2008),
Pedicelli et al.\ (2010), and Storm et al.\ (2011). 
The compilation of Storm et al. (2011) contains 70 Galactic 
fundamental-mode Cepheids, which includes all Cepheids of 
both Fouqu\'e et al.\ (2007) and Groenewegen (2008). 
Therefore, we compare our Cepheid distances with
the results of Storm et al.\ (2011); see Fig.~\ref{fig:dist_comp} (b).
The average distance difference between these two methods is about
1.2\%.
Ngeow (2012) adopted the $V, I$-band Wesenheit function to
derive individual distances to Galactic Cepheids; the comparison of
228 Cepheid distances is shown in Fig.~\ref{fig:dist_comp} (c). The
average distance difference between these two methods is about 0.57\%.
Finally, we compared our distances with open-cluster Cepheid distances
measured by means of the main-sequence fitting method. The latest
results are from Chen et al.\ (2017). For our comparison, see
Fig.~\ref{fig:dist_comp} (d). The average distance difference between
these two methods is only 0.36\%.
In summary, the systematic deviation in the resulting distances
derived here with respect to previously published values is less than
1--2\%.
This confirms that our method is indeed very useful in determining
individual Cepheid distances.

Note that we did not reject any Cepheids with relative large distance
discrepancies when comparing our distances with those published by
other authors. For example, in Fig.~\ref{fig:dist_comp} (a) all data
points lie pretty much on the $y=x$ line, except for FF Aql, which has
$\mu_0$ ({\sl HST}) = 7.76 $\pm$ 0.14 mag and $\mu_0$ (this work) =
8.00 $\pm$ 0.10 mag. FF Aql is the brightest star among the classical
Cepheids thus far observed with {\sl Gaia}. It has a parallax of
$\varpi_{\rm TGAS}$ = 1.640 $\pm$ 0.89 mas, which is consistent with
the object's measured {\sl HST} parallax to within 2$\sigma$. This
star is, in fact, a binary system and its parallax measurements may
therefore be affected by its binary nature. We did not remove it from
our sample. However, some studies, including Fouqu\'e et al. (2007)
and Ngeow (2012), remove outlier Cepheids from their further
discussion. The accuracy of the PL relation depends on the rejection
of outliers (Fouqu\'e et al. 2007). Our near-IR optimal distances
method can be used to determine distances to a few thousand Cepheids,
and more accurate PL relations could thus be achieved.

\section{The Galactic Mid-IR Cepheid Period--Luminosity Relations}

\subsection{The Cepheid Sample with Mid-IR Data}

\begin{figure}[h!]
\centering
\vspace{-0.0in}
\includegraphics[angle=0,width=6.0in]{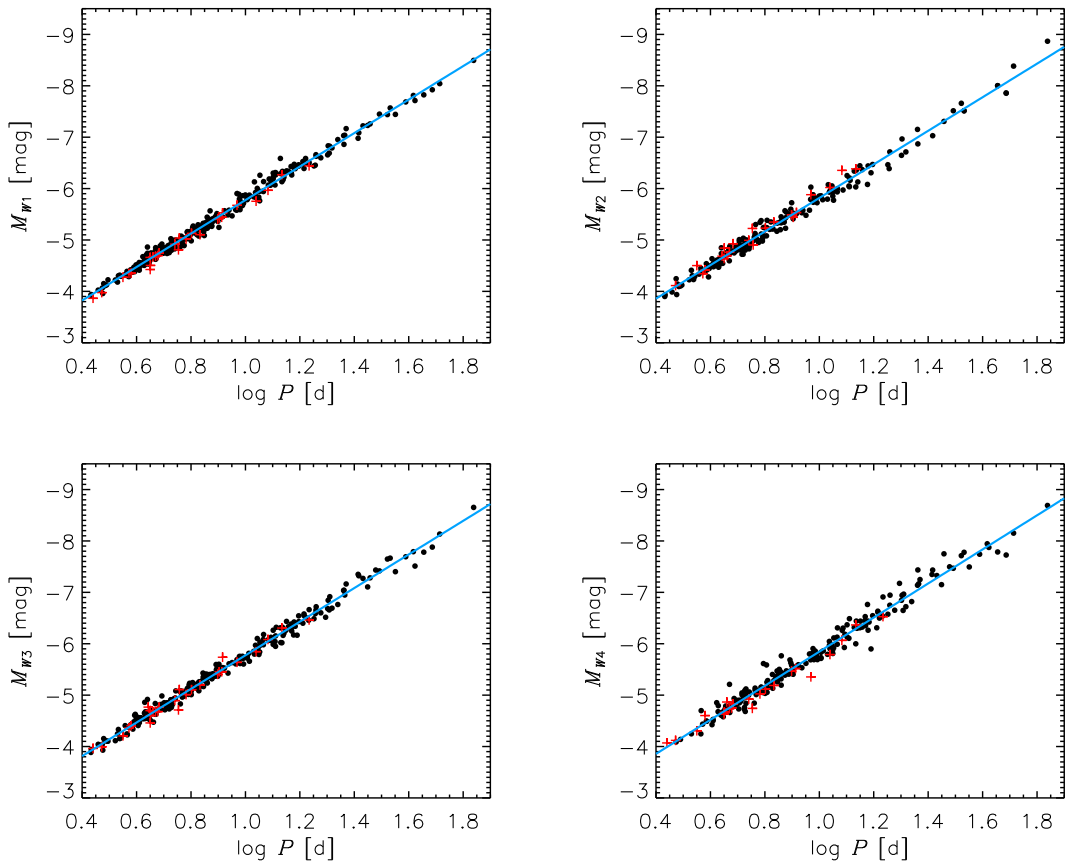}
\vspace{-0.2in}
\caption{\footnotesize
               \label{fig:WISE_PL}
               {\sl WISE} PL relations with distance moduli listed in
               Table 1. The black dots are the
                 fundamental-mode classical Cepheids; the red crosses
                 are the first-overtone classical Cepheids; the blue
                 lines are the best-fitting linear results for all
                 Cepheids, including the black dots and red crosses.}
\end{figure}
%
\begin{figure}[h!]
\centering
\vspace{-0.0in}
\includegraphics[angle=0,width=6.0in]{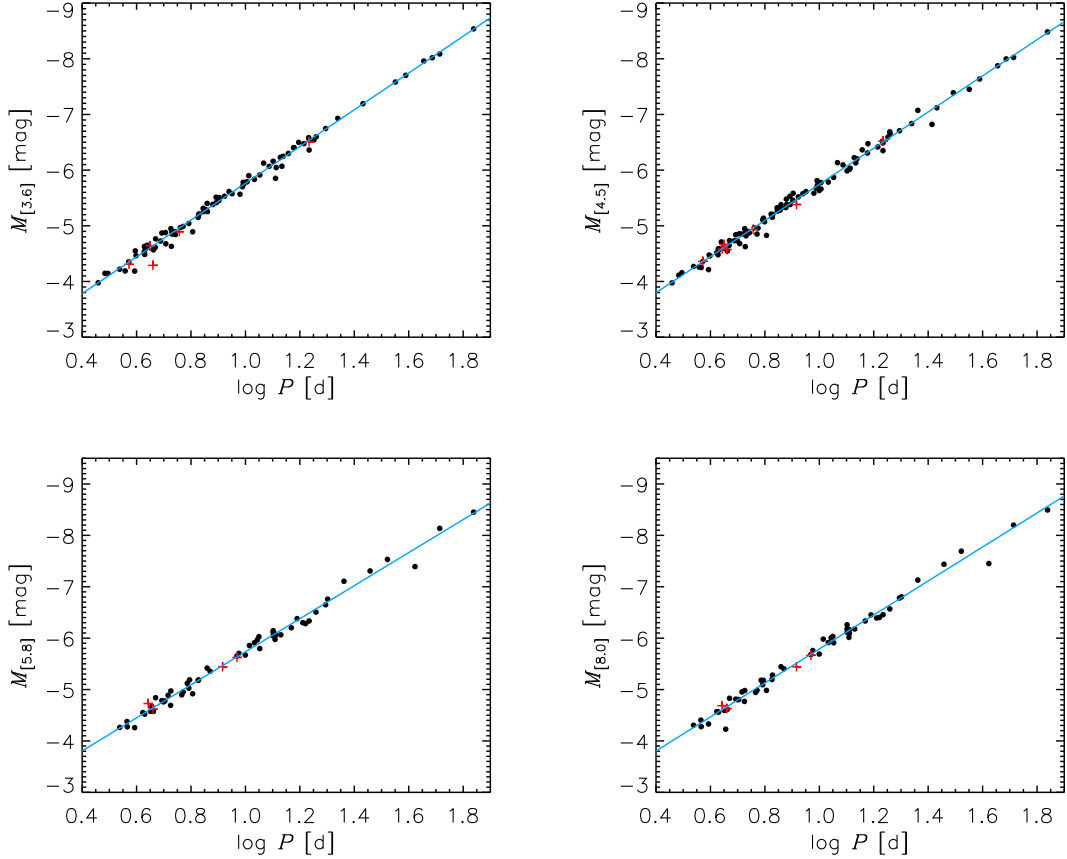}
\vspace{-0.2in}
\caption{\footnotesize
               \label{fig:Spitzer_PL}
               As Fig. 2, but for the {\sl Spitzer} bands. }
\end{figure}

Construction of Galactic mid-IR PL relations requires information
about the distance and extinction to each Cepheid in one's sample. The
availability of distance moduli pertaining to a large sample (288
Galactic classical Cepheids) offers the possibility to determine
accurate mid-IR PL relations. Therefore, we collected the relevant
photometric data from {\sl WISE} and {\sl Spitzer} surveys for our
sample Cepheids.

The {\sl WISE} survey is a full-sky, mid-IR survey with a 40 cm
space-borne telescope (Wright et al.\ 2010). It mapped the sky in the
$W1, W2, W3$, and $W4$ bands (with isophotal central wavelengths of
3.35, 4.60, 11.56, and 22.09 $\mum$, respectively) with 5$\sigma$
limiting magnitudes of about 16.5, 15.5, 11.2, and 7.9 mag,
respectively (Wright et al.\ 2010). We take the {\sl WISE} photometric
data of our Cepheid sample from the AllWISE Multi-epoch Photometry
Database, which provides time-tagged, profile-fit flux measurements for
each object in the AllWISE Source Catalog and Reject Table
\footnote{http://irsa.ipac.caltech.edu/cgi-bin/Gator/nph-dd}. 
The numbers of observations taken for each Cepheid are different,
ranging from 20 to hundreds of visits. For each Cepheid in our sample,
we used contamination and confusion flags (cc\_flags) from the AllWISE
catalog to reject those data points that may be contaminated or biased
in their photometric and/or position measurements. We then calculated
the weighted average values in each band by adopting the reciprocal of
the measurement square error as weights. Finally, the weighted average
values are adopted for the mid-IR $W1, W2, W3$, and $W4$ band mean
magnitudes; these latter measurements are also tabulated in Table 1
(columns 6--9).
  
The Galactic Legacy Infrared Midplane Survey Extraordinaire (GLIMPSE)
program is a mid-IR survey in four bands ([3.6], [4.5], [5.8], and
[8.0]) using the Infrared Array Camera (IRAC) on board the {\sl
  Spitzer} Space Telescope. The isophotal central wavelengths are
3.550, 4.439, 5.731, and 7.872 $\mum$, respectively. The survey data
include {\sl Spitzer} observations from a number of programs covering
the Galactic plane: GLIMPSE I, GLIMPSE II, GLIMPSE 3D, GLIMPSE 360,
Vela-Carina, Deep GLIMPSE, SMOG, and Cygnus-X (Benjamin et al.\ 2003;
Churchwell et al.\ 2009). We search all catalogs for photometric data
of our sample Cepheids. The [3.6], [4.5], [5.8], and [8.0]-band
mean magnitudes
for each Cepheid in our sample are listed in Table 3. In addition,
Monson et al.\ (2012) used 37 Galactic Cepheids with {\sl
  Spitzer}/IRAC [3.6] and [4.5]-band photometric measurements to
calibrate the Galactic Cepheid PL relations. Their sample covers
Cepheid periods ranging from 4 to 70 days. Therefore, we include
their Cepheids to supplement the number of objects with log($P$) $>
1.2$ [days].

\begin{table}[h!]
\begin{center}
\caption{\label{tab:dist} {\sl Spitzer}/IRAC mean magnitudes for
  Galactic Cepheids\tablenotemark{a}}
\vspace{0.1in}
\begin{tabular}{lccccccccc}
\hline \hline
Cepheid & log($P$) & $\langle [3.6] \rangle$  & $\langle [4.5] \rangle$  & $\langle [5.8] \rangle$  & $\langle [8.0] \rangle$  \\ 
  & (days) & (mag) & (mag) & (mag) & (mag) \\
\hline
AN AUR		& 1.012	&	7.033(0.029)	&	--	&	--	&	--\\
ER AUR		& 1.196	&	8.219(0.039)	&	--	&	--	&	--\\
YZ AUR		& 1.260	&	6.538(0.035)	&	6.466(0.019)	&	--	&	--\\
AV TAU		& 0.558	&	8.397(0.035)	&	8.300(0.026)	&	--	&	--\\
AX AUR		& 0.484	&	9.117(0.037)	&	9.134(0.025)	&	--	&	--\\
... & ... & ... & ... & ... & ...\\
\hline
\end{tabular}
\tablenotetext{a}{The entire table is available in the online journal. 
  A portion is shown here for guidance regarding its form and content.}
\end{center}
\normalsize
\end{table}

In summary, we collected eight-band mid-IR mean magnitudes from the
{\sl WISE} and {\sl Spitzer} survey programs. In the $W1, W2, W3$, and
$W4$ bands, the total numbers of Cepheids with available mean
magnitudes are 282, 212, 286, and 219, respectively. In the [3.6],
[4.5], [5.8], and [8.0] bands, the numbers are 90, 106, 59, and 59, respectively.

\subsection{Galactic Mid-IR PL Relations}

\begin{table}[h!]
\begin{center}
\caption{\label{tab:PL} Parameters of the Galactic mid-IR PL Relations}
\vspace{0.1in}
\begin{tabular}{ccccc}
\hline \hline
Band ($\lambda$) & $N$ & $a_\lambda$  & $b_\lambda$ & $\sigma$\\
\hline
\multicolumn{5}{c}{All Cepheids}\\
\hline
$W1$  & 282 & $-3.258\pm0.018$ & $-2.519\pm0.017$ & 0.082\\
$W2$  & 212 & $-3.266\pm0.027$ & $-2.551\pm0.026$ & 0.108\\
$W3$  & 286 & $-3.270\pm0.019$ & $-2.505\pm0.019$ & 0.090\\
$W4$  & 219 & $-3.315\pm0.030$ & $-2.530\pm0.031$ & 0.123\\
$[3.6]$ & 90   & $-3.302\pm0.023$ & $-2.461\pm0.023$ & 0.066\\
$[4.5]$ & 106 & $-3.246\pm0.023$ & $-2.499\pm0.023$ & 0.071\\
$[5.8]$ & 59   & $-3.216\pm0.042$ & $-2.519\pm0.043$ & 0.097\\
$[8.0]$ & 59   & $-3.307\pm0.040$ & $-2.482\pm0.041$ & 0.091\\
\hline
\multicolumn{5}{c}{Excluding First-Overtone Cepheids}\\
\hline
$W1$  & 255 & $-3.248\pm0.018$ & $-2.533\pm0.018$ & 0.082\\
$W2$  & 190 & $-3.266\pm0.027$ & $-2.545\pm0.027$ & 0.107\\
$W3$  & 258 & $-3.263\pm0.020$ & $-2.512\pm0.020$ & 0.090\\
$W4$  & 197 & $-3.317\pm0.032$ & $-2.534\pm0.033$ & 0.125\\
$[3.6]$ & 85   & $-3.298\pm0.024$ & $-2.467\pm0.024$ & 0.066\\
$[4.5]$ & 99  & $-3.245\pm0.024$ & $-2.502\pm0.024$ & 0.073\\
$[5.8]$ & 55   & $-3.222\pm0.044$ & $-2.511\pm0.045$ & 0.099\\
$[8.0]$ & 55   & $-3.311\pm0.042$ & $-2.479\pm0.044$ & 0.093\\
\hline
\hline
\end{tabular}
\end{center}
\end{table}
%

As elaborated by Wang et al.\ (2014), numerous observations appear to
suggest that the mid-IR extinction at $3 \mum < \lambda < 8\mum$ is
$\sim0.5$ times lower than in the near-IR $\Ks$ band. The average
$\Ks$-band extinction, $\langle \AKs \rangle$ for the 288 Cepheids in
our sample is 0.17 mag. This implies that the mid-IR extinction is
about 0.09 mag, which contributes to propagation of $\sim$4\%
uncertainties in distances. Mid-IR extinction corrections should be
considered, although they are usually ignored. Because the mid-IR
extinction is largely independent of the exact sightline compared with
the extinction at shorter wavelengths, we use the average Galactic
extinction from Wang et al.\ (2015) to correct for the mid-IR
extinction in this paper.

With the Cepheid distances derived in Section 2.2 and this extinction correction, 
the absolute magnitudes in the {\sl WISE} and {\sl Spitzer} bands 
can now be calculated for each Cepheid in our sample. 
By means of straightforward linear fits to the log $P$ versus
absolute magnitude diagrams, the mid-IR PL relations are determined. 
Figures~\ref{fig:WISE_PL} and \ref{fig:Spitzer_PL} show the
best-fitting results for the mid-IR {\sl WISE} and {\sl Spitzer} PL relations, respectively. 
The black dots represent the classical Cepheids, 
while the blue solid line is our linear fit. 
The parameters defining our mid-IR multi-band PL relations 
are summarized in Table~\ref{tab:PL} upper panel as ``All Cepheids." 
Overall, the statistical errors in our mid-IR PL relations are small. 
They are less than 0.1 mag except in the $W4$ band (0.12 mag).

  Note that there are 28 first-overtone Cepheids in our sample
  of 288 Galactic Cepheids (Section 2.1). Their $\log P$ ranges from
  0.440 to 1.234 [days], with only three of them characterized by
  $\log P > 1.0$ [days]. Therefore, we also exclude these objects in
  our derivations of the mid-IR {\sl WISE} and {\sl Spitzer} PL
  relations. More specifically, there are 27, 22, 28, and 22 such
  sources in the $W1, W2, W3$, and $W4$ bands, and five, seven, four,
  and four of these objects in $[3.6]$, $[4.5]$, $[5.8]$, $[8.0]$
  bands, respectively. They are indicated as red crosses in
  Figs~\ref{fig:WISE_PL} and \ref{fig:Spitzer_PL}. The parameters of
  the PL relations based on only the fundamental-mode Cepheids are
  listed in Table~\ref{tab:PL} down panel as ``Excluding First-overtone Cepheids". 
  Comparison of these results with those obtained for all Cepheids 
  shows that the slopes and zero points are the same within the uncertainties. 

  We also investigated whether these mid-IR PL relations may
  include any possible nonlinearities. We adopted a nonparametric
  regression technique to explore this aspect. In brief, this method
  allowed us to obtain good fits to the data points without the need
  for a `linear' assumption (for details, see Section 4 of Chen et
  al.\ 2016). For the eight mid-IR bands, we analyzed the differences
  between the nonparametric regression and the linear fit results.
  The differences are small ($<$ 0.03 mag) and exhibit predominantly
  random deviations for all eight bands and for all period ranges,
  which implies that our mid-IR PL relations unlikely contain
  nonlinear features.

\section{Discussion}
\subsection{Uncertainties in the Near-IR Optimal Distances Method}

Independent Cepheid distances can be obtained from trigonometric
parallaxes, the IR surface brightness method, and the main-sequence
fitting method. These distances are usually used to determine and
constrain the zero points of Cepheid PL relations.
Although distances determined based on the near-IR optimal distances
method are indirect distances, by virtue of the large size of our
Cepheid sample (288 sources), the overall distance uncertainties are
small (see Section 2.2). The near-IR optimal distances method depends
on both the near-IR extinction laws and the PL relations. Therefore,
we test if our adopted different near-IR theoretical extinction laws
or the near-IR PL relations could cause systematic differences in
determining Cepheid distances.

\begin{figure}[h!]
\centering
\vspace{-2.7in}
\includegraphics[angle=0,width=5.5in]{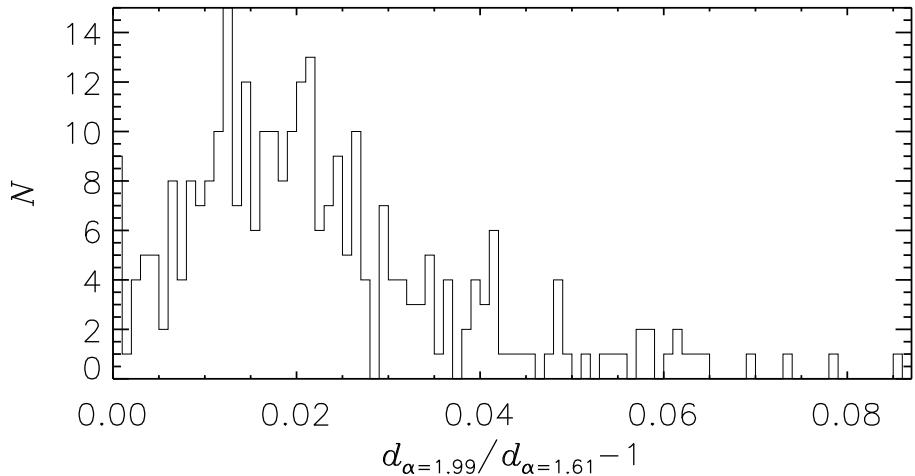}
\vspace{-0.2in}
\caption{\footnotesize
               \label{fig:dist_alpha}
               Comparison of Cepheid distances determined by adopting
               different theoretical extinction laws,
               $A_{\lambda}\propto{\lambda^{-\alpha}}$, with
               $\alpha=1.61$ or 1.99, to correct our Cepheids for the
               effects of near-IR extinction.  }
\end{figure}
%

In Section 2.1, we adopted a power law
$A_{\lambda}\propto{\lambda^{-\alpha}}$ with $\alpha =1.61$ to correct
for the Cepheids' near-IR extinction. A steeper power law, with an
index of $\alpha =1.99$, has also been used to correct for the
extinction in the heavily obscured Galactic bulge. Hence, we discuss
the discrepancies in Cepheid distances caused by adopting this larger
value of $\alpha$. The comparison result is shown in
Fig.~\ref{fig:dist_alpha}. It is apparent that the distances become
systematically larger when adopting $\alpha =1.99$. There are 18
Cepheids (6\% of the total number of 288 Cepheids) with distance
discrepancies in excess of 5\%. These Cepheids do not have any
specific properties compared with the other Cepheids in our
sample. The average distance discrepancy is $\sim 2.2\%$, with a
standard deviation of 1.5\%.
This small systematic error underscores that our distance
determination method is reliable, even when considering variations in
the interstellar environment.

We have adopted the near-IR PL relations of Chen et al.\ (2017), 
which are currently the most complete near-IR Cepheid PL relations, 
based on 31 open-cluster Cepheids with distances determined 
using the main-sequence fitting method. 
In our distance uncertainties analysis (Section 2.2), 
we considered the maximum systematic errors in these PL relations, 
concluding that they contribute $\sim 3.4$\% 
to the uncertainty in the resulting distances (Table 2). 
To investigate the effects of adopting different near-IR PL relations, 
we also use the PL relations of Strom et al.\ (2011). 
The distance discrepancy caused by adopting
different near-IR PL relations is $\sim 1.4$\%, with a standard
deviation of 1.4\%, which is less than the systematic error of 3.4\%. 
This means that the published near-IR PL relations agree well
with each other given the prevailing uncertainties, and adopting 
different PL relations does not measurably affect our distance determination.

\subsection{Advantages of Galactic Mid-IR PL Relations}

The experiential four-band {\sl WISE} classical Cepheid PL relations
were determined for the first time based on our sample of hundreds of
Cepheids (see Table 4). These PL relations are characterized by high
accuracies. The uncertainties in the mid-IR PL relations are 0.08,
0.1, 0.09, and 0.12 mag in the $W1, W2, W3$, and $W4$ bands,
respectively. Compared with the uncertainties in the current near-IR
PL relations (e.g., 0.155, 0.146, 0.144 mag in the $J, H, \Ks$ bands:
Fouqu\'e et al.\ 2007; 0.22 mag in the $J, K$ bands: Strom et
al.\ 2011; 0.148, 0.124, 0.120 mag in the $J, H, \Ks$ bands: Chen et
al.\ 2017; 0.155, 0.146, 0.144 mag in the $J, H, K$ bands: Madore et
al.\ 2017), these mid-IR uncertainties are even smaller. This means
that more accurate distances to classical Cepheids could be obtained
based on these {\sl WISE} PL relations. The small dispersions of $<
0.12$ mag in the {\sl WISE} PL relations also underscore the accuracy
of the {\sl WISE} photometry of bright classical Cepheids. As the
full-sky {\sl WISE} survey provides multi-epoch photometric data,
these PL relations could be used to determine distances to a few
thousand Cepheids.

Our {\sl Spitzer} PL relations are based on the largest Cepheid sample
available to date: it is composed of triple the number of objects
compared with the 29 Cepheids of Ngeow (2012) and twice the number of
sources compared with the 37 objects of Monson et al.\ (2012) with
observations in the [3.6] and [4.5] bands. This represents a
significant improvement. Compared with previous determinations, the
newly derived {\sl Spitzer} PL relations agree well with previously
published results based on other methods, given the associated
uncertainties, and the accuracy is significantly improved. With
respect to the slopes determined by Ngeow (2012), we find differences
of 0.060 and 0.066 mag dex$^{-1}$ in the [3.6] and [4.5] bands. While
our slope of the [3.6] PL relation is consistent (discrepancy: 0.008
mag dex$^{-1}$) with that of Monson et al.\ (2012), who fixed the
slope to $-3.31$ and determined their [3.6]-band PL relation using
{\sl Spitzer} Large Magellanic Cloud data. They calibrated the zero
point of the [3.6] PL relation at $-5.80 \pm 0.03$ mag by relying on
the geometric {\sl HST} guide-star distances to 10 Galactic
Cepheids. The discrepancy in the {\sl Spitzer} [3.6]-band PL
relation's zero point between our derivation and that of Monson et
al.\ (2012) is only 0.019 mag.

We also compared the {\sl WISE} PL relations with the {\sl Spitzer} PL
relations. As the isophotal wavelengths of the $W1$ and $W2$ bands
are comparable with those of the [3.6] and [4.5] bands, the slopes of
our PL relations in these sets of bands are comparable. Among the
mid-IR PL relations in the eight bands available, the {\sl Spitzer}
[3.6]-band PL relation has the lowest uncertainty (0.066 mag), which
propagates to $\sim$3\% uncertainties in the resulting distances.
This uncertainty is smaller than the uncertainties in the Galactic
near-IR PL relations (e.g., 0.22 mag: Storm et al.\ 2011; $>$ 0.12
mag: Chen et al.\ 2017), and even smaller than the uncertainties in
the LMC PL relations ($>$ 0.11 mag in the [3.6] and [4.5] bands:
Scowcroft et al.\ 2011; $>$ 0.09 mag in the $J, H, \Ks$ bands: Macri
et al.\ 2015). It is also smaller than the uncertainties in
  the SMC PL relations ($>$ 0.16 mag in the IRAC bands: Ngeow \&
  Kanbur 2010; $>$ 0.21 mag in the {\it AKARI} $3.2\mum$ and $4.1\mum$
  bands: Ngeow et al.\ 2012). 
Moreover, our uncertainty of 0.066 mag provides an upper limit to the
width of the Cepheid instability strip.

\subsection{Mid-IR Extinction}
%

\begin{figure}[h!]
\centering
\vspace{-0.0in}
\includegraphics[angle=0,width=16.0in]{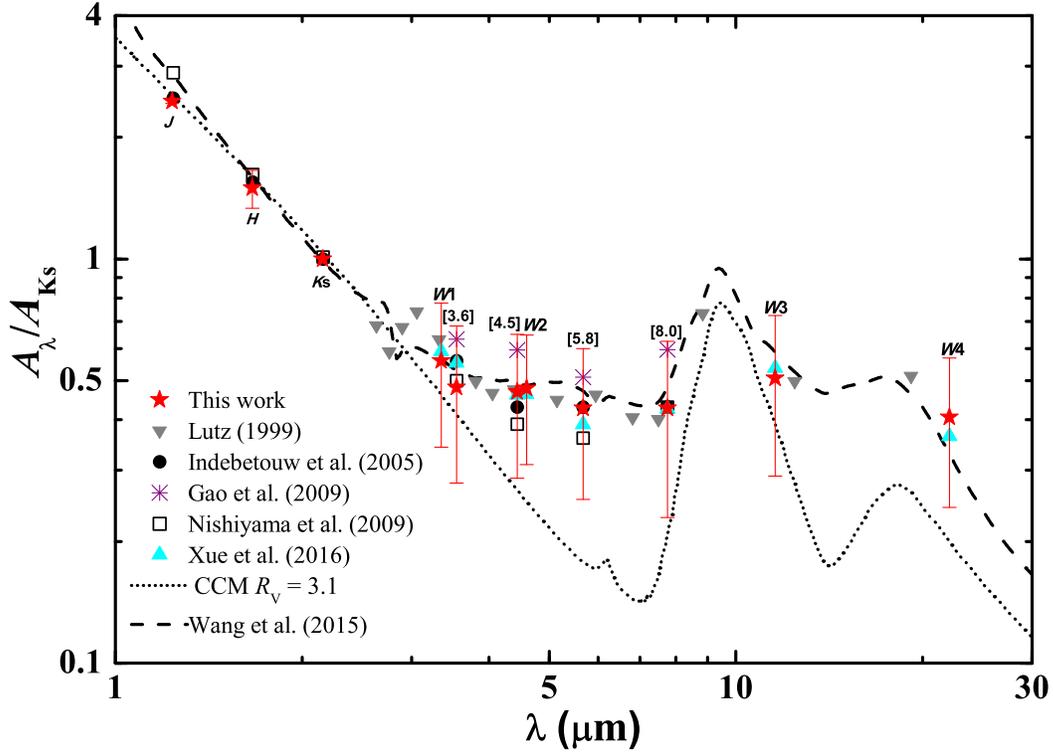}
\vspace{-7.2in}
\caption{\footnotesize
               \label{fig:mid_ext}
               Comparison of the extinction derived in this paper
               (red stars) with previous determinations (different
               symbols). The CCM $\Rv=3.1$ model and the Wang et
               al.\ (2015) ice model extinction curves are also shown.
}
\end{figure}

Using the mid-IR intensity-averaged magnitudes and multi-band PL
relations, we calculate the mid-IR extinction for individual
Cepheids. The mean mid-IR extinction values (relative to $\AKs$) in
the four {\sl WISE} bands and the four {\sl Spitzer} bands are 
$A_{W1}/A_\Ks \approx 0.560\pm0.218$, $A_{W2}/A_\Ks \approx
0.479\pm0.169$, $A_{W3}/A_\Ks \approx 0.507\pm0.217$, $A_{W4}/A_\Ks
\approx 0.406\pm0.163$, $A_{[3.6]}/A_\Ks \approx 0.481\pm0.202$,
$A_{[4.5]}/A_\Ks \approx 0.469\pm0.182$, $A_{[5.8]}/A_\Ks \approx
0.427\pm0.173$, and $A_{[8.0]}/A_\Ks \approx 0.427\pm0.198$ mag. The
results are illustrated in Fig.~\ref{fig:mid_ext} as red stars.
Previous determinations for other lines of sight (e.g., Lutz 1999;
Indebetouw et al.\ 2005; Gao et al.\ 2009; Nishiyama et al.\ 2009; Xue et al.\ 2016) 
are represented by different symbols. For comparison, the CCM $\Rv=3.1$
model and the Wang et al.\ (2015) ice model extinction curves are also
shown.

\section{Summary}

We have introduced a near-IR optimal distances method to determine the
distances to Galactic classical Cepheids. Based on these newly derived
distances, the mid-IR PL relations have been tightly constrained. The
major results of this paper are as follows:
\begin{enumerate}
\item Distances to the overall sample of 288 Galactic classical
  Cepheids have been determined. The global uncertainty is less than
  4.9\%.
\item Comparison of our distance moduli with those from literature
  sources based on {\sl HST} parallaxes, the IR surface brightness
  method, Wesenheit functions, and the main-sequence fitting
  method. The average systematic discrepancy between our results and
  sets of published distances is less than 1--2\%.
\item Galactic mid-IR PL relations for Cepheids in the four {\sl WISE}
  bands ($W$1, $W$2, $W$3, and $W$4) have been constructed for the
  first time, based on a sample containing hundreds of Cepheids. PL
  relations in the four {\sl Spitzer}/IRAC bands ([3.6], [4.5], [5.8]
  and [8.0]) have also been constructed, resulting in significant
  improvements in the associated uncertainties. Among the published PL
  relations, our {\sl Spitzer} [3.6]-band PL relation has the smallest
  dispersion 0.066 mag.
\item The mean mid-IR extinction curve for Cepheids has been obtained:
  $A_{W1}/A_\Ks \approx 0.560$, $A_{W2}/A_\Ks \approx 0.479$,
  $A_{W3}/A_\Ks \approx 0.507$, $A_{W4}/A_\Ks \approx 0.406$,
  $A_{[3.6]}/A_\Ks \approx 0.481$, $A_{[4.5]}/A_\Ks \approx 0.469$,
  $A_{[5.8]}/A_\Ks \approx 0.427$, and $A_{[8.0]}/A_\Ks \approx 0.427$
  mag.
\item Finally, based on our sample of 91 Cepheids with [3.6]-band
  absolute magnitudes, the [3.6]-band PL relation zero point of
  Freedman et al.\ (2012) can be well constrained. If we adopt the
  slope of $-3.31$ pertaining to the LMC Cepheids, the systematic
  error in the distance modulus to the LMC is reduced from 0.033 mag
  to $0.066/\sqrt{91}=0.007$ mag. Combining the observational data for
  80 LMC Cepheids, a distance modulus of 
  $\mu_0 = 18.457 \pm 0.011 \mbox{(statistical)} \pm 0.007$ (systematic) mag 
  is obtained. Consequently, the uncertainty in the absolute zero point
  of the PL relation decreases from 1.7\% to 0.3\%, and this also
  means that the zero point of the [3.6]-band PL relation is no longer
  a major contributor to remaining uncertainty in the Hubble constant.

\end{enumerate}

\acknowledgments{We thank Noriyuki Matsunaga for very helpful
  discussions. This work is partially supported by the Initiative
  Postdocs Support Program (No.\ BX201600002) and the China
  Postdoctoral Science Foundation 2017M610998.  S.W. acknowledges
  support from a KIAA Fellowship. R.dG. is grateful for funding
  support from the National Natural Science Foundation of China
  through grants U1631102, 11373010, and 11633005.}


\end{document}